\documentclass[aps,prb,twocolumn,superscriptaddress,showpacs]{revtex4-1}
\usepackage{epstopdf}
\usepackage{epsfig}
\usepackage{amsmath}
\usepackage{amssymb}
\usepackage[colorlinks=true,citecolor=blue,linkcolor=blue]{hyperref}
\newcommand{\ket}[1]{\left|#1\right>}

\newcommand{\braket}[1]{\left<#1\right>}
\newcommand{\para}[1]{\left(#1\right)}
\newcommand{\abs}[1]{\left|#1\right|}

\newcommand{\sd}[0]{\ \ \ }

\begin{document}

\title{Slave-spin approach to the strongly correlated systems}

\author{Masoud Mardani}
\affiliation{Department of Physics, Sharif University of Technology, Tehran 11155-9161, Iran}

\author{Mohammad-Sadegh Vaezi}
\affiliation{Department of Physics, Sharif University of Technology, Tehran 11155-9161, Iran}

\author{Abolhassan Vaezi}
\email{Corresponding author: vaezi@ipm.ir}
\affiliation{School of Physics, Institute for Research in Fundamental Sciences, IPM, Tehran, 19395-5531, Iran}

%\author{S. Akbar Jafari}
%\affiliation{Department of Physics, Sharif University of Technology, Tehran 11155-9161, Iran}
%\affiliation{School of Physics, Institute for Research in Fundamental Sciences, IPM, Tehran, 19395-5531, Iran}

\begin{abstract}
In this paper, we develop a new type of slave particle method which is similar to the slave rotor model except that the quantum rotor is substituted by a spin one slave particle. The spin-one slave particle itself can be represented in terms of Schwinger bosons/fermions. This approach is more conveniently applicable to the strongly correlated Hamiltonians with onsite Hubbard interaction and resolves the limitations of using the slave rotor model as well as the Anderson-Zou slave particle technique. For instance, the mean-field parameters of the slave spin method do not vanish above the Mott transition and this approach is smoothly connected to the non-interacting limit. As an example, we study the phase diagram of the Kane-Mele-Hubbard model using our current approach. In the absence of the spin-orbit interaction, the Mott transition occurs at $U_c\simeq 3 t_1$. Several aspects of the slave spin method, its gauge theory and various possible mean-field states associated with this approach have been discussed.
\end{abstract}
%\date{\today}

%\pacs{71.10.Fd,71.10.Pm,73.20.-r}

%71.10.Fd 	Lattice fermion models (Hubbard model, etc.)
%72.80.Ga Transition-metal compounds, electrical conductivity of
% 03.65.Vf Topological phases (quantum mechanics)
%71.10.Pm Fermions in reduced dimensions (anyons, composite fermions, Luttinger liquid, etc.)
%73.20.-r 	Electron states at surfaces and interfaces

\maketitle
\section{Introduction}

\begin{figure}[tbp]
\begin{center}
\includegraphics[width=200pt]{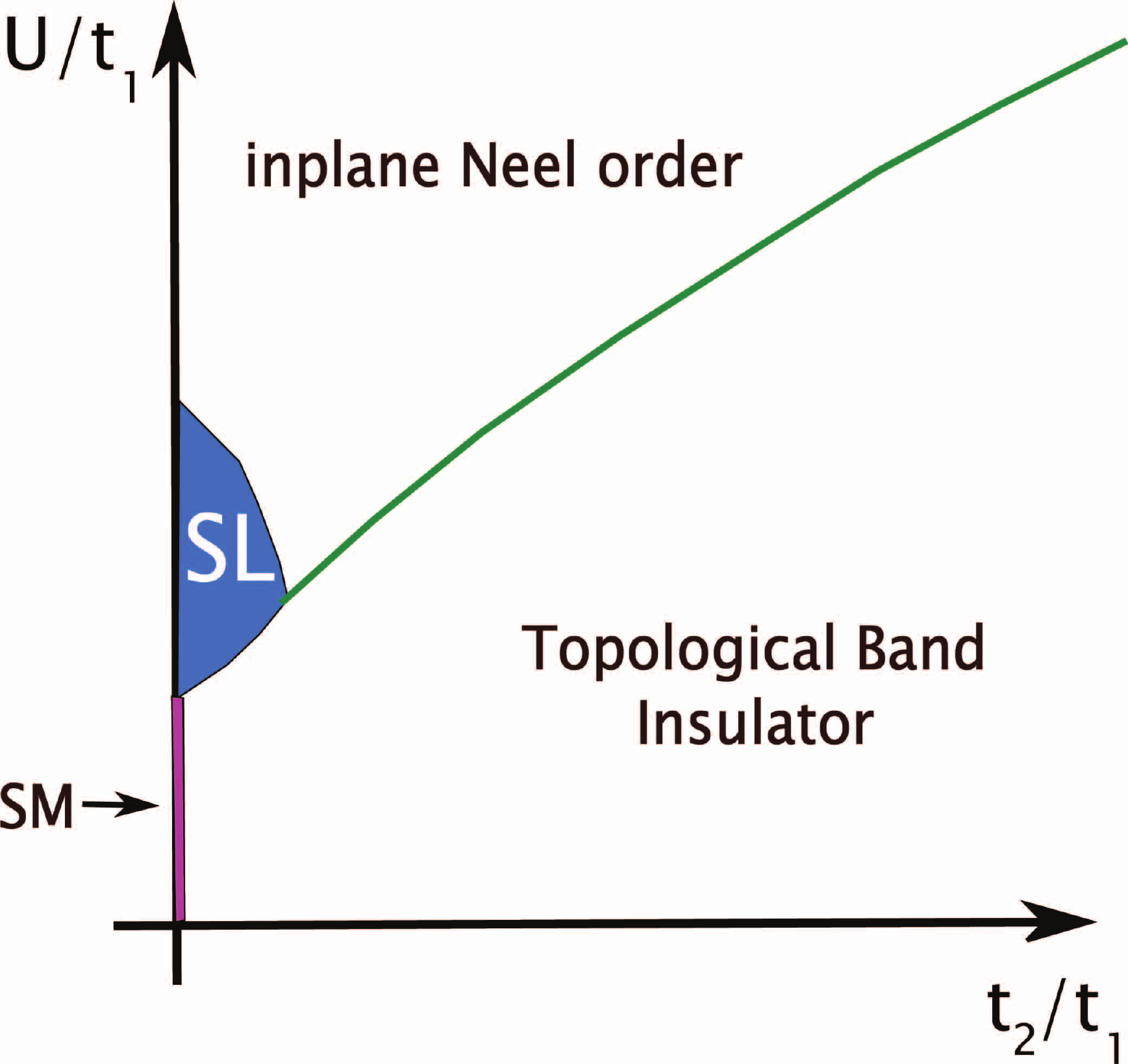}
\caption{(Color online) Schematic phase diagram of the slave spin approach for the KMH model. $t_2$ is the spin-orbit coupling constant. There are four phases we can identify in the slave spin model: 1- Semi-metal (SM) phase, where both charge and spin excitations are gapless. 2- Gapped spin liquid (SL) phase which according to our theory is going to be a chiral spin liquid state. In this phase both charge and spin degrees of freedom are gapped. 3- In-plane Neel ordering. In this phase we initially obtained a gapless spin liquid where charge is gapped but spin remains gapless. The spin liquid was coupled to a gapless and compact $U(1)$ gauge field. After including the instanton effect, the spin liquid phase undergoes a phase transition into the in-plane (XY) Neel ordering. 4- Topological band insulator (TBI). In this phase charge gap closes, but spin degree of freedom is gapped in the bulk of the system. The spin up and spin down spinons have opposite Chern numbers and therefore, it corresponds to an insulator with gapless helical edge states.}\label{Fig0}
\end{center}
\end{figure}

The theoretical proposal for the topological insulators (TI) and their experimental observations have drawn a huge interest in the past few years \cite{Zhang_Qi_2010_a,Hasan_2010_1,Bernevig_2006_1,Hsieh_2008_1}.
This new state of matter hosts several interesting phenomena. For instance, TIs  have topologically protected helical edge states and exhibit spin quantum Hall effect accordingly. Most of theoretical studies in this field have been carried out on the weakly interacting systems. However, the effects of interaction on the properties of TIs remain largely unclear. The intuition is that they are robust against weak interactions. However, exotic states of matter or spontaneously broken phases may emerge due to strong inetraction/corelation. Therefore, it deems necessary to study the interacting TI (ITI) in order to have a better insight into the physics of TIs. The simplest interaction that one can consider is the onsite Hubbard interaction. Recently, several authors have studied the phase diagram of the Kane-Mele-Hubbard (KMH) model \cite{Kane_Mele_2005_1,Kane_Mele_2005_2} both theoretically \cite{LeHur_2010_a,vaezi_2011_0116,LeHur_2011_a,Varney_2011_a,Fiete_2011_b,vaezi_2011_d,Yu_2011_a,Yamaji_2010_a} and numerically \cite{Wu_2010_1,Assad_2010_a,Assaad,Heidarian_2011_1}.

In this paper, the issue of onsite Hubbard interaction in the electronic systems has been addressed. We develop a new {\em slave-spin} technique that is suitable to study a number of famous models in the strongly correlated systems e.g. Anderson, Hubbard and the KMH models \cite{Slave_rotor_2,Meng_2010a,Mosadeq_2010_a,Schmidt_2010_1}. It is worth mentioning that the slave-spin model was first introduced by de' Medici et. al. in Ref. \cite{Medici_2005_1} and was studied further in Refs. \cite{Medici_2008_1,Ruegg_2009_1}. However, our slave-spin model is different than the previously known slave-spin model, though both approaches share the same physical idea. The main difference is that in the previous studies, every electron was attached to its own {\em spin 1/2} slave-spin, while in our approach, both spin up and spin down electrons are attached to the same {\em spin one} slave-spin. Additionally, the main point of our present work is to represent the slave-spin in terms of Schwinger bosons or Schwinger fermions. This is a very crucial feature of our model as the spin operators are hard to work with. In the previous slave-spin studies, authors did not employ this decomposition to deal with the slave spins.

In the present work, our slave spin approach has been applied to the last two models. This approach has two important advantages over the two other well-known slave particle approaches, the Anderson-Zou \cite{Vaezi_2010a} and the slave rotor models \cite{Slave_Rotor_1,Senthil_2008_1}. Firstly, it is smoothly connected to the non-interacting case and has all the desired properties. This was missed in the Anderson-Zou slave particle technique at half filling. For example, instead of a semi-metal phase, Anderson-Zou slave approach gives a superconducting phase for the Hubbard model on the honeycomb lattice. Secondly, our model can be easily applied above the Mott transition and provides the correct behavior in the large $U/t_1$ limit. However, the slave rotor cannot be easily applied in that limit as the mean-field parameters vanishes in the simplest approximation.

\section{Method}
Electron carries both spin and electric charge. The spin up and spin down electrons carry the same electric charge but opposite spins. Similar to the slave rotor model, the electron operator in the slave spin model is decomposed into a charge creating and a spin creating parts. This is a natural way to implement the idea of spin-charge separation in the deconfined phases. Therefore, we have

\begin{eqnarray}
  c_{i,\sigma}^\dag = S^{+}_{i}f_{i,\sigma}^\dag ,
\end{eqnarray}
where $S^{+}_{i}$ creates the charge and $f_{i,\sigma}^\dag$ the spin of electron. The $f_{i,\sigma}^\dag$ slave particles that carry the spin of electrons are usually referred to as spinons. Since the charge at site $i$, has three possible values $0,-e,-2e$, we assign the charge to the $S^{z}_{i}=n_{e}(i)-1$ eigenvalues of a spin one object, where $n_e(i)$ is the total number of electrons at site $i$. Doubly occupied sites correspond to $S^z=+1$, half filled sites carry $S^z=0$ and empty site correspond to $S^z=-1$. Subsequently, $S^{+}_{i}$ and $S^{-}_{i}$ are the ladder operators for the spin one object corresponding to the charge of electrons that connects different charge sectors. The physical spin of doubly occupied sites as well as empty sites is zero, however the spin of $S^z=0$ sector can be either $+1/2$ or $-1/2$ corresponding to spin up and spin down electrons. The following table summarizes the relations

The electron operator in terms of new slave particles needs to be project in order to recover the physical Hilbert space. This can be done by implementing the following constraint on the physical Hilbert space.

\begin{eqnarray}
  S^z_{i}=n_{f}\para{i}-1=f_{i,\uparrow}^\dag f_{i,\uparrow}+f_{i,\downarrow}^\dag f_{i,\downarrow}-1.
\end{eqnarray}

As a sanity check, the above constraints yields four sites for the dimension of the local Hilbert space. Another way to confirm this result is to consider the gauge freedom in the definition of slave particles. It is clear from Eq. [1] that the electron operator is invariant under the following U(1) gauge transformation

\begin{eqnarray}
  && f_{i,\sigma}^\dag\to \exp\para{i\theta_i} f_{i,\sigma}^\dag \sd,\sd S^{\pm}_{i}\to \exp\para{\mp i\theta_i} S^{\pm}_{i}.~~~
\end{eqnarray}

That is the both spinons and $S^{+}$ carry opposite charges under the internal U(1). On the other hand, $S^{+}$ ($S^{-}$) carries $+1$ (-1) charge under the spin rotation group along $z$ axis ($S^z$ charge), because it changes the eigenvalue of $S^z$ operator by +1 (-1). Due to the gauge invariance of the physical Hilbert states, the total charge of spinons at site $i$ has to be equal to the $S^z_i$ up to a shift which can be fixed by acting $c_{i,\sigma}^\dag$ on one physical state.

Now, let us use Schwinger slave boson (fermion) representation for the slave spin operators, $S^{\pm,z}_{i}$. Because the Hilbert space of the slave spin has three states, we can assign a hardcore boson (fermion) with three flavors to each charge state (see Table II). To be more precise, $b_{1}$ flavor is assigned to the empty site, $b_{2}$ flavor to the singly occupied and $b_{3}$ flavor to the doubly occupied sites. Therefore, the eigenvalue of the $S^z_i$ operator is the difference between the number of flavor three and one Schwinger bosons (fermions). Also, the $S^+_i$ operator acts on the states by changing the flavor of Schwinger bosons(fermions). The following equations summarizes the scheme

\begin{table}
\begin{tabular}{|c|c|c|c|}
  \hline
  %physical state &state in terms of $c$ operators  & state in terms of slave operators & $n^f$ & $S^z$ \\
  Physical states &  states in terms of slave operators  & $n^f_i$ & $S^z_i$ \\ \hline
   $\ket{0}_i$&$ \ket{0}\ket{-1}_c $& 0 & -1 \\ \hline
  $\ket{\uparrow}_i $&$ f_{i,\uparrow}^\dag\ket{0}_s S_i^{+}\ket{-1}_c$& 1 & 0 \\ \hline
  $\ket{\downarrow}_i $&$ f_{i,\downarrow}^\dag\ket{0}_s S_i^{+}\ket{-1}_c$& 1 & 0 \\ \hline
  $\ket{\uparrow \downarrow}_i $&$f_{i,\uparrow}^\dag f_{i,\downarrow}^\dag \ket{0}_s S_i^{+}\para{S_i^{+}\ket{-1}_c}$& 2 & 1 \\
  \hline
\end{tabular}
\caption{The local physical Hilbert space that satisfies the $S^z_{i}=n_{f}\para{i}-1$ constraint.}
\end{table}

\begin{eqnarray}
 && S_{i}^{+}=b_{3,i}^\dag b_{2,i}+b_{2,i}^\dag b_{1,i},\sd S_{i}^{-}=b_{2,i}^\dag b_{3,i}+b_{1,i}^\dag b_{2,i}\cr
 && S_{i}^{z}=b_{3,i}^\dag b_{3,i}-b_{1,i}^\dag b_{1,i}.
\end{eqnarray}

From now on, we only consider Schwinger bosons in calculations. The Schwinger boson representation enjoys the following internal U(1) gauge degree of freedom (different from that was discussed earlier).

\begin{eqnarray}
&&b_{i,m}^\dag \to \exp\para{i\alpha_i}b_{i,m} ^\dag \sd,\sd m=1,2,3
\end{eqnarray}
Therefore they all carry the same charge under the above internal U(1) gauge field and is equal to their number density. Therefore the total number of Schwinger fermions has to be fixed. Another way to obtain this result is to consider the dimension of the Hilbert space of the charge sector which is three. To achieve that result, the number of Schwinger bosons has to be fixed and equal to one. Therefore,

\begin{eqnarray}
 && b_{1,i}^\dag b_{1,i}+b_{2,i}^\dag b_{2,i}+b_{3,i}^\dag b_{3,i}=1.
\end{eqnarray}

In the path integral formalism, the constraints in equations [2] and [6] can be implemented using two Lagrange multiplier fields and integrating over those auxiliary fields.

\begin{table}
\begin{tabular}{|c|c|c|c|}
  \hline
  %physical state &state in terms of $c$ operators  & state in terms of slave operators & $n^f$ & $S^z$ \\
   $S^z$& charge states &  states in terms of slave bosons  & $n^b$  \\ \hline
  -1& $\ket{-1}_c$&$ \ket{1,0,0}=b_{1,i}^\dag\ket{0,0,0}_b $& 1  \\ \hline
  0& $\ket{0}_c=S^+_i\ket{-1}_c $&$ \ket{0,1,0}=b_{2,i}^\dag\ket{0,0,0}_b$& 1\\ \hline
  +1&$\ket{+1}_c=S^+_i\ket{0}_c $&$\ket{0,0,1}=b_{3,i}^\dag\ket{0,0,0}_b$& 1 \\ \hline
\end{tabular}
\caption{The local physical Hilbert space of the charge sector that satisfies the $n^b_{i}=n^b_{1,i}+n^b_{2,i}+n^b_{3,i}=1$ constraint.}
\end{table}

\section{Onsite Hubbard interaction}
The Hubbard interaction acts locally and adds a cost to the creation of doubly occupied sites through $Un_{i,\uparrow}n_{i,\downarrow}$ term where $n_{i,\sigma}$ is the number of spin $\sigma$ electron at site $i$. Since $n_{i,\uparrow}n_{i,\downarrow}$ counts the number of doubly occupied sites, i.e. the states with $S^z_{i}=n^f-1=1$, it is equal to the number of $b_{3,i}$ Schwinger bosons. Therefore $n_{i,\uparrow}n_{i,\downarrow}=b_{3,i}^\dag b_{3,i}$. So we have

\begin{eqnarray}
 && \sum_{i} n_{i,\uparrow}n_{i,\downarrow}=\sum_{i} b_{3,i}^\dag b_{3,i}
\end{eqnarray}

On the other hand, due to the $S^z_i=n^f_i-1$ constraint, at half filling we have
\begin{eqnarray}
  \sum_{i}S^z_{i}=\sum_i \para{b_{3,i}^\dag b_{3,i}-b_{1,i}^\dag b_{1,i}}=\sum_{i} \para{n^{f}_i-1}=0.~~
\end{eqnarray}
So we can rewrite the Hubbard interaction as follows

\begin{eqnarray}
 && H_{\rm U}=\frac{U}{2}\sum_{i}\para{b_{3,i}^\dag b_{3,i}+b_{1,i}^\dag b_{1,i}}.
\end{eqnarray}

\section{Gauge theory of the slave spin model}
As we discussed before, there two U(1) internal gauge degrees of freedom. $U(1)_{s}$ represents the gauge the gauge freedom in the definition of electron operators and $U(1)_{c}$ that represents the gauge freedom in the Schwinger boson representation of spin slaves. Table III summarizes the transformation properties and the charge of different slave particles under $U(1)_{c}$ and $U(1)_{s}$.

\begin{table}
\begin{tabular}{|c|c|c|c|}
  \hline
  %physical state &state in terms of $c$ operators  & state in terms of slave operators & $n^f$ & $S^z$ \\
   slave operator& $U(1)_{EM}$ charge& $U(1)_{s}$ charge& $U(1)_{c}$ charge   \\ \hline
  $f_{i,\uparrow}^\dag$& $-q_{s}$&1 & 0  \\ \hline
  $f_{i,\downarrow}^\dag$& $-q_{s}$&1&0\\ \hline
  $b_{i,1}^\dag$&$1-q_{s}$&-1&1 \\ \hline
  $b_{i,2}^\dag$&$0$&0&1 \\ \hline
  $b_{i,3}^\dag$&$-1+q_{s}$&1&1 \\ \hline
\end{tabular}
\caption{The charge of the different slave operators under electromagnetic $U(1)_{EM}$, $U(1)_{s}$ and $U(1)_{c}$ gauge fields. The transformation of charge $q$ operators after $\theta_{i}$ gauge transformation is $\hat{O}_{q}(i)\to  \exp\para{i\theta_i}\hat{O}_{q}(i)$. The transformation of $b_{i,m}^\dag$ quasiparticles under $U(1)_{s}$ have been chosen so that $S^{\pm}_i\to \exp\para{\mp i \theta_{i}}S^{\pm}_i$ and $S^z_i \to S^z_i$. The electromagnetic charge of spinons $-q_{s}$ can be assumed to be any number. This ambiguity will be resolved if we appropriately take the $U(1)_{s}$ gauge field into consideration which leads to the generalized Ioffe-Larkin formula. For more details see Refs. \cite{Vaezi_2011_3311} and \cite{Vaezi_SC} .}
\end{table}

Due to the compact nature of $U(1)_{s}$ and $U(1)_{c}$ gauge fields, both fields have to be gapped to obtain a stable and physical mean-field results. Otherwise, instanton proliferation destabilize the mean-field state. Gauge fields can be gapped out in two ways: 1. through Anderson-Higgs mechanism, 2. by the presence of Chern-Simons action in the low energy physics. In the following, we comment on the possible outcomes of the slave spin model.

\subsection{$U(1)_{s}$ gapped, $U_{c}$ gapped}
In this case, gauge field is gapped and instanton operator is irrelevant. Therefore, the mean-field calculations can be trusted. Four possibilities may happen:

{\bf 1- Both spin and charge degrees of freedom are gapless.---} The spin gap can be attributed to the gap in the spectrum of spinons ($f_{i,\sigma}$ particles). The charge gap on the other hand is physically related to the doublon (doubly occupied site) or holon (empty site) creation. Hence, the charge gap can be attributed to the gap in the spectrum of $b_{3,i}$ and $b_{1,i}$ Schwinger bosons (note that $b_{2,i}$ represent the singly occupied sites). When instantons are absent and both spin and charge degrees of freedom are gapless, i.e. there is no gap in the spectrum of spinons and $b_{3,i}$ or $b_{1,i}$ slave particles, the mean-field state is nothing but a conducting phase and is a Fermi liquid (FL) system.

{\bf 2- Spin is gapped while charge is gapless.---} This phase corresponds to the band insulator, where band-structure calculations for electrons predict a filled band that is separated from the conduction band by a nonzero gap. Although the above statement is valid for the bulk of the system, there may be gapless edge states present. Two famous examples are the non-interacting Kane-Mele and Haldane models, where they host helical and chiral edge states respectively, though their bulk are insulating.

{\bf 3- Spin remains gapless while charge is gapped.---} This phase corresponds to the Mott insulating phase, where band-structure calculations for electrons predict a metallic phase, while taking correlations into account changes the behavior of the system. In Mott insulator, charge cannot be excited due to the nonzero charge gap while spin degree of freedom may or may not be gapless. If there is no magnetic ordering in the ground-state of a Mott insulator, the system is called a spin liquid. In this case, we obtain a gapless spin liquid.

{\bf 4- Both spin and charge are gapped.---} This phase simply corresponds to the gapped spin liquid provided that there is no magnetic ordering in the ground-state.

\subsection{$U(1)_{s}$ gapped, $U(1)_{c}$ gapless}
In this case, $U(1)_{c}$ gauge field is gapless and instanton operator that adds a quantum of flux of the corresponding field through quantum tunneling is most likely a relevant perturbation. Under instanton proliferation, the mean-field state becomes unstable and we cannot trust our results. Instanton derives the system into the confinement phase where only $U(1)_{c}$ gauge invariant operators can be measured and any other operator vanishes.

\subsection{$U(1)_{s}$ gapless, $U(1)_{c}$ gapped}
In this case, $U(1)_{s}$ gauge field is gapless and again instantons will proliferate and the mean-field result cannot be trusted.

\subsection{$U(1)_{s}$ gapped, $U(1)_{c}$ gapless}
In this case, both internal gauge fields have gapless modes that results in instanton proliferation. The mean-field state is again unstable and slave particle is not a good approach in this case.

\section{Phase diagram of the Kane-Mele-Hubbard model}

The Kane-Mele-Hubbard model on the honeycomb lattice is described as
\begin{equation}
\label{KMH-model}
H=-t_{1} \sum_{\langle ij \rangle,\sigma} c^{\dag}_{i,\sigma}c_{j,\sigma}+U\sum_{i} n_{i\uparrow}n_{i\downarrow} + it_{2}\sum_{\langle\langle i,j \rangle\rangle,\sigma}\sigma\nu_{ij}c^{\dag}_{i,\sigma}c_{j,\sigma}
\end{equation}
where $t_{1}$, $U$, and $t_{2}$ are the nearest neighbor hopping energy, the strength of the on-site repulsion, and  the second-neighbor spin-orbit coupling strength, respectively. Here $c_{i\sigma}$ ($c_{i\sigma}^{\dag}$) annihilates (creates) an electron with spin $\sigma$ on site $i$. $\nu_{i,j}$ is introduced so as to obtain a nonzero flux turning around any triangular path and is defined as $\nu_{i,j}=\frac{\vec{d}_{i}\times \vec{d}_{j}}{\abs{\vec{d}_{i}\times \vec{d}_{j}}}. \hat{z}$ where $d_{i}$ and $d_j$ are two shortest vectors that connect sites $i$ and $j$ i.e. $\vec{d}_i+\vec{d}_j=\vec{R}_i-\vec{R}_j$ (see Fig. [2]).

\begin{figure}[tbp]
\begin{center}
\includegraphics[width=120pt]{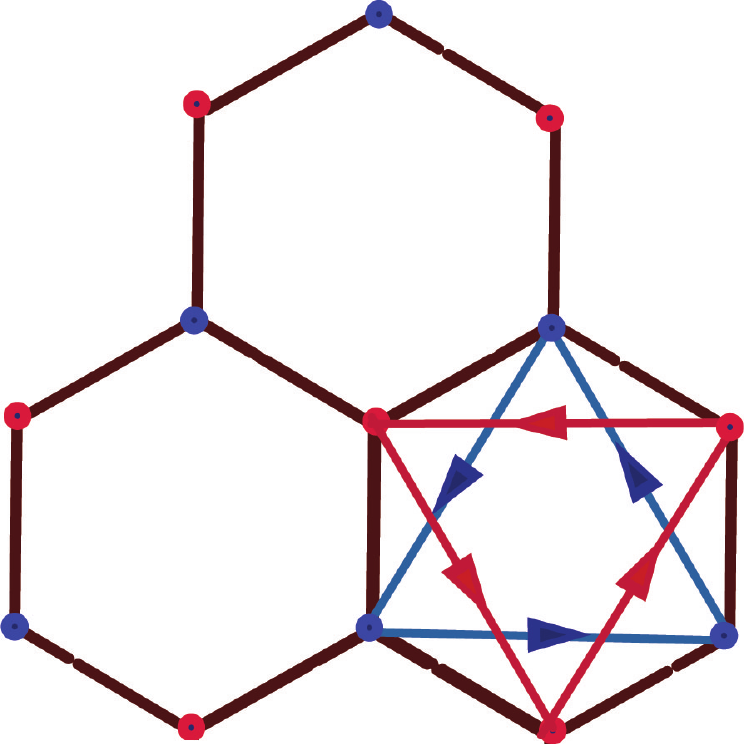}
\caption{(Color online) Kane-Mele model for the topological insulators. Arrows denote the phase of the hopping to the NNN for the spin up electrons. The phase for the spin downs is opposite to that of spin up electrons. Assigning a phase $\phi_{\sigma}$ to the hopping of spin $\sigma$ electrons induces a $3\phi_{\sigma}$ magnetic flux for spin $\sigma$ electrons moving around the depicted triangles. It is this nonzero spin dependent flux that causes spin quantum Hall effect for non-interacting electrons.}\label{Fig2}
\end{center}
\end{figure}

Now we can employ the slave spin model to rewrite the KMH model in terms of spinons and Schwinger fermions. To do so, note that
\begin{eqnarray}
&& c_{i,\sigma}^\dag c_{j,\sigma}= S^{+}_{i}S^{-}_{j}f_{i,\sigma}^\dag f_{j,\sigma}.
\end{eqnarray}
where $S^{+}_{i}$ ($S^{-}_i=S^{+~\dag}_i$) is defined in terms of Schwinger bosons in Eq. [4]. Now we can decouple spinons from slave spins using the following Hartree-Fock approximation that can be justified through Hubbard-Stratonovic transformation

\begin{eqnarray}
 c_{i,\sigma}^\dag c_{j,\sigma} \simeq && \braket{S^{+}_{i}S^{-}_{j}}f_{i,\sigma}^\dag f_{j,\sigma}+S^{+}_{i}S^{-}_{j}\braket{f_{i,\sigma}^\dag f_{j,\sigma}}\cr
&& -\braket{S^{+}_{i}S^{-}_{j}}\braket{f_{i,\sigma}^\dag f_{j,\sigma}}.
\end{eqnarray}

Now let us define the $\chi_{\sigma}\para{i,j}$ mean-field parameter in the following way

\begin{eqnarray}
&&\chi_{\sigma}\para{i,j}=\braket{f_{i,\sigma}^\dag f_{j,\sigma}}.
\end{eqnarray}
For the nearest neighbor, we assume a uniform and spin independent $\chi_{1}\para{i,j}$ as follows
\begin{eqnarray}
&&\chi^{1}_{\sigma}\para{i,j}=\chi_{1}/2.
\end{eqnarray}
For the second nearest neighbor, except in the chiral spin liquid phase that will be discussed later, we assume that it takes the following value
\begin{eqnarray}
&&\chi^{2}_{\sigma}\para{i,j}=i\nu_{i,j}\sigma\chi_{2}/2.
\end{eqnarray}
Using the definition of the slave spin operators we have
\begin{eqnarray}
 &&S^{+}_{i}S^{-}_{j}=\para{ b_{3,i}^\dag b_{2,i}+b_{2,i}^\dag b_{1,i}}\para{ b_{2,j}^\dag b_{3,j}+b_{1,j}^\dag b_{2,j}}.
\end{eqnarray}
In the Schwinger boson approach, it is more convenient to assume $\braket{b_{m,i}^\dag b_{n,j}}=0$ where $n,m=1,2,3$. Consequently,
\begin{eqnarray}
 &&S^{+}_{i}S^{-}_{j}\simeq \Delta_{22}\para{i,j} b_{3,i}^\dag b_{1,j}^\dag+\Delta_{13}\para{i,j}^{*} b^\dag_{2,i} b^\dag_{2,j} \cr
&&+H.c.-2Re\para{\Delta_{22}\para{i,j}\Delta_{13}^{*}\para{i,j}},
\end{eqnarray}
where
\begin{eqnarray}
\Delta_{13}\para{i,j}=\braket{b_{1,i}b_{3,j}},
\end{eqnarray}
and
\begin{eqnarray}
\Delta_{22}\para{i,j}=\braket{b_{2,i}b_{2,j}}.
\end{eqnarray}
As a result,
\begin{eqnarray}
\braket{S^{+}_{i}S^{-}_{j}}=2Re\para{\Delta_{22}\para{i,j}\Delta_{13}^{*}\para{i,j}}.
\end{eqnarray}
We assume $\Delta_{22}\para{i,j}$ and $\Delta_{13}\para{i,j}$ are real and independent of direction and only depend on the distance of two neighbors. The Hubbard term can also be written as in Eq. [9]. The microscopic constraints on the Hilbert space of slave particles in Eqs. [2] and [6] can also be implemented at the mean-field level using Lagrange multipliers $\lambda_1$ and $\lambda2$. It is straightforward that at half filling, $\lambda_1=0$ naturally satisfies the constraint in average. Using all these procedures, we finally obtain the following form for the effective mean-field Hamiltonian
\begin{eqnarray}
  H_{eff}=H_{f}+H_{b}+H_{\rm cl.}
\end{eqnarray}
where

\begin{eqnarray}
\label{S-F-Matrix-1}
&&H_{f}=-\sum_{k,\sigma} \para{\begin{array}{cc}
                           f_{k,A,\sigma}^\dag & f_{k,B,\sigma}^\dag
                         \end{array}} \left(
                                                                                                               \begin{array}{cc}
                                                                                                                  t^*_2\sigma\zeta_k& t^*_1\eta_{k} \\
                                                                                                                 t^*_1\eta_{k}^{*} & -t^*_2\sigma\zeta_k \\
                                                                                                               \end{array}
                                                                                                             \right)
                         \para{\begin{array}{c}
                            f_{k,A,\sigma} \\
                            f_{k,B,\sigma}
                          \end{array}}\cr
\end{eqnarray}
where
\begin{eqnarray}
\label{S-F-Matrix-2}
&&t^*_1=2t_1\Delta^{1}_{22}\Delta^{1}_{13}\\
&&t^*_2=2t_2\Delta^{2}_{22}\Delta^{2}_{13}.
\end{eqnarray}
and the following structure factors have been used
\begin{eqnarray}
&&\eta_{k}=\exp\para{-ik_y}+2\cos\para{\sqrt{3}k_x/2}\exp\para{ik_y/2}\\
&&\zeta_{k}=2\sin\para{\sqrt{3}k_x/2}\para{\cos\para{3k_y/2}-\cos\para{\sqrt{3}k_x/2}}\sd
\end{eqnarray}
Thus, the energy spectrum of the spinons are
\begin{eqnarray}
  E^{\pm}_{f,k}=\pm \sqrt{\para{t^*_2\sigma\zeta_k}^2+\abs{t^*_1\eta_{k}}^2}
\end{eqnarray}
At half filling, all the negative energy levels are occupied and the gap in the spectrum is proportional to $\abs{t_2}$. Although the bulk is insulating, the edge of the system can be shown through various techniques e.g. Jackiw-Rebbi solitons \cite{Jackiw_Rebbi} host gapless spinon excitations, even in the disordered system. The number of gapless modes for each spin equals to the $\abs{C_{\sigma}}$, where $C_{\sigma}$ is the first Chern number of the band structure. Using the continuum limit of the above Hamiltonian that yields two gapped Dirac cones can be used to compute the Chern index which is going to be $C_{\uparrow}=-C_{\downarrow}=C$ and $\abs{C}=1$.

Now let us comment on the effective Hamiltonian for Schwinger bosons which is
\begin{eqnarray}
H_{b}=&&\para{\frac{U}{2}+\lambda_2}\sum_{i}\para{b_{3,i}^\dag b_{3,i}+b_{1,i}^\dag b_{1,i}}\cr
&&-2t_1\chi_{1} \sum_{<i,j>}\Delta^{1}_{22}\para{b_{3,i}^\dag b_{1,j}^\dag+b_{3,j}b_{1,i}} \cr
&&-2t_1\chi_{1} \sum_{<i,j>}\Delta^{1}_{13}\para{b_{2,i}^\dag b_{2,j}^\dag+b_{2,j}b_{2,i}} \cr
&&-2t_2\chi_{2} \sum_{<<i,j>>}\Delta^{2}_{22}\para{b_{3,i}^\dag b_{1,j}^\dag+b_{3,j}b_{1,i}} \cr
&&-2t_2\chi_{2} \sum_{<<i,j>>}\Delta^{2}_{13}\para{b_{2,i}^\dag b_{2,j}^\dag+b_{2,j}b_{2,i}}
%&&+N_{s}\para{12t_1\chi_{1}\Delta^{1}_{22}\Delta^{1}_{13}+24 t_2   \chi_{2}\Delta^{2}_{22}\Delta^{2}_{13}+\lambda_2}\cr
\end{eqnarray}
Since $b_{3}$, and $b_{1}$ Schwinger bosons are paired and are decoupled from $b_{2}$ Schwinger bosons at the mean-field level, we have
\begin{eqnarray}
&&H_{b}=H_{b}^{13}+H_{b}^{22},
\end{eqnarray}
in which
\begin{eqnarray}
&&H_{b}^{13}=\sum_{k} \Phi_{13,k}^\dag M_{k}^{13} \Phi_{13,k}.
\end{eqnarray}
We have used the following definitions
\begin{eqnarray}
&&\Phi_{13,k}^\dag=\para{b_{1,A,k}^\dag, b_{1,B,k}^\dag, b_{3,A,-k}, b_{3,B,-k}}.
\end{eqnarray}
and
\begin{widetext}
\begin{eqnarray}
&&M_{k}^{13}=\left( \begin{array}{cccc}
          U/2-\lambda_2 & 0 & -t_2\chi_2\Delta^{2}_{22}\xi_k & -t_1\chi_1\Delta^{1}_{22}\eta_{k} \\
          0 & U/2-\lambda_2 & -t_1\chi_1\Delta^{1}_{22}\eta_{k}^{*} & -t_2\chi_2\Delta^{2}_{22}\xi_k \\
          -t_2\chi_2\Delta^{2}_{22}\xi_k & -t_1\chi_1\Delta^{1}_{22}\eta_{k} & U/2-\lambda_2 & 0 \\
          -t_1\chi_1\Delta^{1}_{22}\eta_{k}^{*} & -t_2\chi_2\Delta^{2}_{22}\xi_k & 0 & U/2-\lambda_2 \\
        \end{array}
      \right).
\end{eqnarray}
where
\begin{eqnarray}
  &&\xi_{k}=2\cos\para{\sqrt{3}k_x/2}\cos\para{3k_y/2}+\cos\para{\sqrt{3}k_x}.\sd
\end{eqnarray}

The energy excitation of $b_3-b_1$ branch can be easily computed using the Bogoliubov transformation and is given by \\

\begin{eqnarray}
&&E_{13,k}^{\pm}=\sqrt{\para{U/2-\lambda_2}^2\pm 2t_1t_2\chi_1\chi_2\Delta^{1}_{22}\Delta^{2}_{22}\abs{\eta_{k}}\xi_{k}-\abs{t_1\chi_1\Delta^{1}_{22}\eta_k}^2-\para{t_2\chi_2\Delta^{2}_{22}\xi_{s,k}}^2}.\cr
&&
\end{eqnarray}

%\end{widetext}

For the $b_2$ branch we have the following pairing Hamiltonian
\begin{eqnarray}
&&H_{b}^{22}=\frac{1}{2}\sum_{k} \Phi_{22,k}^\dag M_{k}^{22} \Phi_{13,k}\cr
&&\Phi_{22,k}^\dag=\para{b_{2,A,k}^\dag, b_{2,B,k}^\dag, b_{2,A,-k}, b_{2,B,-k}}\cr
&&M_{k}^{22}=-\left(
        \begin{array}{cccc}
          \lambda_2 & 0 & 2t_2\chi_2\Delta^{2}_{13,k}\xi_k & 2t_1\chi_1\Delta^{1}_{13}\eta_{k} \\
          0 & \lambda_2 & 2t_1\chi_1\Delta^{1}_{13}\eta_{k}^{*} & 2t_2\chi_2\Delta^{2}_{13}\xi_k \\
          2t_2\chi_2\Delta^{2}_{13}\xi_k & 2t_1\chi_1\Delta^{1}_{13}\eta_{k} & \lambda_2 & 0 \\
          2t_1\chi_1\Delta^{1}_{13}\eta_{k}^{*} & 2t_2\chi_2\Delta^{2}_{13}\xi_k & 0 & \lambda_2 \\
        \end{array}
      \right).
\end{eqnarray}
The energy spectrum of the above Hamiltonian can also be computed through the Bogoliubov transformation and is given by following expression
\begin{eqnarray}
  &&E_{22,k}^{\pm}=\sqrt{\lambda_2^2\pm 8t_1t_2\chi_1\chi_2\Delta^{1}_{13}\Delta^{2}_{13}\abs{\eta_{k}}\xi_{k}-\abs{2t_1\chi_1\Delta^{1}_{13}\eta_k}^2-\para{2t_2\chi_2\Delta^{2}_{13}\xi_{s,k}}^2}.\cr
  &&
\end{eqnarray}
\end{widetext}

There is also a classical contribution to the energy which can be calculated by adding c-numbers in the procedure
\begin{eqnarray}
\frac{H_{\rm cl.}}{N_s}=12t_1\chi_{1}\Delta^{1}_{22}\Delta^{1}_{13}+24 t_2   \chi_{2}\Delta^{2}_{22}\Delta^{2}_{13}+\frac{5}{2}\lambda_2-\frac{U}{2}.~~~
\end{eqnarray}
where $N_{s}$ is the number of sites. To obtain the variational mean-field parameters, we need to compute the total ground-state energy and minimize it with respect to its parameters. It is given by the following relation

\begin{eqnarray}
E_{\rm g}=&&\sum_{k}\para{E^{+}_{13,k}+E^{-}_{13,k}+\frac{E^{+}_{22,k}+E^{-}_{22,k}}{2}-2\abs{E^{f}_{k}}}+H_{\rm cl.}\cr
&&
\end{eqnarray}

\section{Possible phases in the Kane-Mele-Hubbard model}

To interpret the results of the slave spin approach to the KMH model, we would like to make several comments

{\bf I-Single condensation versus pair condensation of bosons}. Boson can condense both as local objects (singly) or form pairs and condense as extended objects. Usually start from pair condensation scenario and solve the Hamiltonian. When the energy excitation gap closes, it is easy to argue that single condensation scenario is energetically more favorable \cite{vaezi_2011_0116}. In the slave spin approach to the Kane-Mele-Hubbard, our numerical study shows that the energy spectrum for the $b_{2}$ Schwinger bosons is gapless in the whole phase diagram and therefore we have

\begin{eqnarray}
  \braket{b_{2,i}^\dag}\neq 0~.
\end{eqnarray}

{\bf II-When do internal gauge fields are gapped?} To gap out a gauge field, we have two options. 1- condense a charged operator. Any operator $\hat{O}_q(x)$ that carries charge $q$ under that gauge field, transforms after $f(x)$ gauge transformation in the following way
\begin{eqnarray}
  \hat{O}_{q}(x)\to \exp\para{if(x)}  \hat{O}_{q}(x).
\end{eqnarray}
If the system is gauge invariant, the expectation value of all physical quantities has to be gauge independent as well, i.e.

\begin{eqnarray}
  \braket{\hat{O}_{q}(x)}=\braket{\exp\para{if(x)}  \hat{O}_{q}(x)}.
\end{eqnarray}

The above equation has two solutions: $q=0$ or $\braket{\hat{O}_{q}(x)}=0$. Now what if we find a charged operator whose expectation value is nonzero? It simply means that the gauge symmetry is broken. This is the reason why Cooper pair formation leads to the Meissner effect and as a result to the superconductivity. When gauge symmetry breaks, gauge particles acquire mass through $m^2 a^{\mu}a_{\mu}$ term in their action. It is easy to show (e.g. by using the Landau-Ginzberg theory of phase transition) that $m^2 \propto \braket{\hat{O}_q}$.

2- Through Chern-Simons term. When the ground-state breaks the time reversal and exhibits quantum Hall effect in response to the gauge field that we are arguing about, the low energy action for that gauge field is given by the following Chern-Simons action
\begin{eqnarray}
  S_{\rm G.F.}=\int d^2x dt ~\frac{C}{4\pi}\epsilon^{\mu\nu\lambda}a_{\mu}\partial_{\nu}a_{\lambda}-\frac{1}{2{\rm e}^2}\para{\epsilon^{\mu\nu\rho}\partial_{\nu}a_{\lambda}}^2\sd.
\end{eqnarray}
It is straightforward to see (e.g. by solving the classical equations of motion for the gauge field) that the gauge particles acquire a nonzero mass proportional to $\abs{C}$.

As we mentioned above, our numerical study for the KMH shows that always $<b_{2,i}^\dag>\neq 0$. Since $b_{2,i}^\dag$ carries nonzero charge under the internal $U(1)_{c}$, its condensation leads to the gauge symmetry breaking of the $U(1)_c$ gauge field. We only need to be concerned about the $U(1)_{s}$ gauge field. Our numerical and theoretical observation at half filling, and when $t_2=0$ is as follows:

{\bf $U<U_{c,1}\simeq 3t_1$ .---} The energy excitation for the $b_{3}$ and $b_{1}$ vanishes below $U_{c,1}$ and they undergo Bose-Einstein condensation. Therefore we have

\begin{eqnarray}
&& t_2=0,~~~ U<U_{c,1}: ~~~ \abs{\braket{b_{3,i}^\dag}}=\abs{\braket{b_{1,i}^\dag}}\neq 0.
\end{eqnarray}
The above equation simply means that the $U(1)_{s}$ gauge symmetry breaks as well below $U_{c,1}$. At $t_2=0$, spinons are gapless as well, and altogether we conclude the system is in the semi-metal phase.

{\bf $U_{c,1}<U<U_{c,2}$ :  $b_{3,1}$ Schwinger bosons are gapped, however spinons form chiral state with nonzero total Chern number.---} This result needs more explanation and is beyond a simple meanfield result. When the charge gap is nonzero ($b_{3}$ and $b_{1}$ Schwinger bosons form bound-state), the charge degree of freedom freezes and can be safely integrated out to find effective action for the spinon degree of freedom. In our case, it corresponds to the loop corrections to the spinon-spinon interaction. Such approach has been studied by one of us in Ref. \cite{vaezi_2011_0116} in detail. When $t_2=0$, the effective Hamiltonian for the spinons is given by $J_1-J_2$ Heisenberg model. In Ref. \cite{vaezi_2011_0116} authors have considered the more general case where spin-orbit interaction is taken into consideration as well and we obtained the so called Kane-Mele-Heisenberg model. They have obtained a region above Mott transition where hosts spinons with gapped chiral ground-sate and nonzero Hall response (nonzero Chern number). The nonzero Chern number of spinons in response to the $U(1)_{s}$ gauge field opens up a nonzero gap in the spectrum of gauge particles and breaks the gauge symmetry. Therefore in this phase, both internal gauge fields are broken and we do not need to worry about instanton effect accordingly. Since, both charge and spin gaps are nonzero and spinons form a chiral state, we call this state {\em gapped chiral spin liquid} which extends to nonzero $t_2$ as well.

{\bf $U>U_{c,2}$ .---} In this case, charge gap is again nonzero, while $J_2$ is very small and is not enough to form chiral state for spinons. Therefore, spinons are gapless in this case. Since non of the Anderson-Higgs mechanism, nor the Chern-Simons action break the $U(1)_s$ gauge field, instantons will proliferate. Instanton effect results in spontaneously broken phases such as Neel order or valence solid bond (VBS) orders \cite{Vaezi_2011_a,vaezi_2011_0116}.

\section{Conclusion} In this paper, we introduced slave-spin model which is similar to the slave rotor model. It was discussed that using this technique is more conveniently applicable as it is applicable in the whole phase diagram and is smoothly connected to the non-interacting electron systems. The gauge theory of this model has been presented. It was argued that this approach explains the existence of gapped spin liquid in the phase diagram in a narrow region for moderate values of $U_c/t_1$ and for small values of $t_2$.

%\bibliography{Refs}
%\bibliography{Slave_Spin}

%

\end{document}